\documentclass{article}
\usepackage{spconf,amsmath,graphicx}
\usepackage{booktabs}
\usepackage{subcaption}
\usepackage{colortbl}
\usepackage{xcolor}
\usepackage{multirow}
\usepackage{amsfonts}

\title{MMW: Side Talk Rejection Multi-Microphone Whisper on Smart Glasses}

\name{Yang Liu, Li Wan, Yiteng Huang, Yong Xu, yangyang shi, Saurabh Adya, ming sun, Florian Metze}
\address{Meta, US}
\begin{document}
%
\maketitle
\begin{abstract}

Smart glasses are increasingly positioned as the next-generation interface for ubiquitous access to large language models (LLMs). Nevertheless, achieving reliable interaction in real-world noisy environments remains a major challenge, particularly due to interference from side speech. In this work, we introduce a novel side-talk rejection multi-microphone Whisper (MMW) framework for smart glasses, incorporating three key innovations. First, we propose a Mix Block based on a Tri-Mamba architecture to effectively fuse multi-channel audio at the raw waveform level, while maintaining compatibility with streaming processing. Second, we design a Frame Diarization Mamba Layer to enhance frame-level side-talk suppression, facilitating more efficient fine-tuning of Whisper models. Third, we employ a Multi-Scale Group Relative Policy Optimization (GRPO) strategy to jointly optimize frame-level and utterance-level side speech suppression. Experimental evaluations demonstrate that the proposed MMW system can reduce the word error rate (WER) by 4.95\% in noisy conditions.
    
\end{abstract}
\section{Introduction}
\label{sec:intro}
Large language models have become central to a wide range of AI applications, including dialogue systems, content generation, and multimodal reasoning \cite{achiam2023gpt, touvron2023llama}. Their integration into wearable devices, such as smart glasses, has opened new opportunities for ubiquitous, voice-based interaction. These devices promise to serve as natural language interfaces for real-time assistance, information retrieval, and task execution, enabling users to access AI systems hands-free and on the move. However, ensuring robust communication between users and LLMs in real-world acoustic environments remains a major challenge.

A particularly pressing issue is side speech interference—unintended speech from nearby individuals or overlapping conversations—which degrades the quality of user-LLM interaction. Unlike traditional close-talk systems, smart glasses rely on open-field microphones at outdoor environment, rendering them especially vulnerable to ambient noise and interfering speakers. While techniques such as wake-word detection, beamforming, and source separation have been proposed, they often fall short in dynamic and unstructured acoustic environments. 


One line of work addresses this challenge by extracting spatial information and fusing it with encoded audio features before passing the signals to an LLM. This approach offers modularity and scalability—only the lightweight spatial encoder needs to be adapted to different hardware platforms, while the LLM itself remains unchanged. Classical signal processing techniques, such as beamforming \cite{van1988beamforming}, generalized cross-correlation (GCC) \cite{knapp2003generalized}, and subspace-based methods like MUSIC \cite{schmidt1986multiple} and ESPRIT \cite{roy1989esprit}, have been extensively studied for estimating the direction of arrival (DoA) from inter-microphone time and phase differences. However, these methods often require large microphone arrays and tend to perform poorly in noisy and reverberant conditions.

To overcome these limitations, recent research has increasingly focused on deep learning-based spatial modeling. Convolutional neural networks (CNNs) \cite{adavanne2018sound}, recurrent neural networks (RNNs) \cite{xu2017convolutional}, and transformer-based models \cite{park2021many} have shown promise in learning spatial features directly from waveforms or spectrograms. For example, DOAnet \cite{adavanne2018sound} leverages a hybrid CNN-RNN architecture to estimate DoAs for multiple sound sources under challenging acoustic environments. Nonetheless, most of these models are trained primarily on environmental or non-speech sounds and depend on extensive microphone arrays, making them unsuitable for compact wearable devices. Recent work \cite{zheng2024bat, tang2024can} has begun to explore perceptual spatial modeling in the context of ambient sound understanding including physically informed acoustic simulation tools~\cite{chen2022soundspaces} and spatially-aware learning algorithms~\cite{yang2022srp, qiao2024joint, devnani2024learning}, but speech-focused implementations remain limited.


Another emerging direction involves fine-tuning LLMs with spatial information, enabling them to selectively attend to audio content from specific directions. While models such as SALMONN \cite{tang2023salmonn} and GAMA \cite{ghosh2024gama} support speech-based tasks in multimodal contexts, they lack explicit spatial grounding, limiting their effectiveness in complex auditory scenes. A major drawback of this approach is the need to train and maintain distinct LLM instances for different hardware configurations. Even with parameter-efficient methods such as LoRA \cite{hu2022lora}, this leads to substantial operational cost and complexity.

Amid these developments, Whisper has emerged as a powerful foundation model for automatic speech recognition (ASR), trained on 680,000 hours of diverse audio data, with half of it in non-English languages \cite{radford2023robust}. Its encoder-decoder architecture supports multilingual transcription, language identification, and translation, demonstrating strong robustness across a wide range of noise conditions and accents. Owing to its generalization ability and open availability, Whisper has become a popular base for building multi-speaker and spatial-aware ASR systems. Recent works have extended Whisper to target-speaker ASR (TS-ASR), enabling it to focus on a designated speaker in multi-talker scenarios. For example, Ma et al. \cite{ma2024extending} explored prompt tuning approaches to adapt Whisper for TS-ASR, while Meng et al. \cite{meng2024empowering} demonstrated a joint model for multi-talker and target-talker recognition. Polok et al. proposed DICoW \cite{polok2024dicow}, a diarization-conditioned Whisper system that incorporates speaker embeddings to improve speaker attribution accuracy. Further, Polok et al. \cite{polok2025target} systematically studied how Whisper can be adapted for TS-ASR without extensive architectural modifications. These approaches highlight Whisper’s flexibility as a backbone for multi-speaker and spatial speech modeling. However, existing methods often depend heavily on diarization or external conditioning modules, and the integration of Whisper with LLMs for spatially-aware natural language interaction remains underexplored.

In this paper, we propose an end-to-end approach for side-talk rejection using a multi-microphone Whisper (MMW) system. In contrast to prior methods, our system neither relies on explicit speaker embeddings or solely depends on frame-level diarization. Instead, it introduces three key innovations: First, A Mix Block based on a Tri-Mamba architecture \cite{liu2024masv} that effectively fuses multi-channel audio at the raw waveform level while preserving compatibility with streaming processing; Second. A Frame Diarization Mamba Layer that enhances frame-level side-talk suppression, facilitating more effective fine-tuning of Whisper models. Spacial information is used for diarizing the speakers. Last, a Multi-Scale Group Relative Policy Optimization strategy that jointly optimizes frame-level and utterance-level side speech suppression.


\section{Related Work}
\label{sec:related_work}

\subsection{Whisper}

OpenAI's Whisper is a powerful end-to-end automatic speech recognition (ASR) system based on a Transformer encoder-decoder architecture, trained through large-scale weak supervision on multilingual audio datasets~\cite{radford2023robust}. Whisper is available in multiple model sizes, including tiny (39M parameters), base (74M parameters), small (244M parameters), medium (769M parameters), and large (1550M parameters), catering to a range of computational budgets and application scenarios.

Several Whisper variants have emerged to enhance its performance or adapt it to specialized tasks. Fine-tuning Whisper on domain-specific or low-resource language data has been widely explored~\cite{liu2024exploration}. Moreover, multi-input variants that integrate multiple microphone signals have demonstrated significant improvements in accuracy for multi-speaker and noisy environments~\cite{meng2024empowering}. In addition, some studies have extended Whisper with video or textual inputs to address multimodal speech recognition and understanding tasks, effectively leveraging complementary cross-modal information~\cite{rouditchenko2024whisper}.

\subsection{Mamba}

State Space Models (SSMs) are designed to map one-dimensional sequences $x(t) \mapsto y(t) \in \mathbb{R}$ through a linear ordinary differential equation (ODE):

\begin{align}
x'(t) &= \mathbf{A}x(t) + \mathbf{B}u(t), \\
y(t) &= \mathbf{C}x(t),
\end{align}
where $\mathbf{A} \in \mathbb{R}^{N \times N}$ is the state matrix, $\mathbf{B}$ and $\mathbf{C} \in \mathbb{R}^N$ are model parameters, and $x(t) \in \mathbb{R}^N$ represents the latent state. Structured State Space Sequence Models (S4)~\cite{gu2021efficiently} improve upon basic SSMs by introducing a structured parameterization of $\mathbf{A}$ and an efficient computation algorithm. The state matrix is initialized using the High-Order Polynomial Projection Operator (HIPPO) framework~\cite{gu2020hippo}, enabling deep sequence models with robust and efficient long-range reasoning capabilities. S4 has been shown to outperform Transformers~\cite{vaswani2017attention} on the Long Range Arena Benchmark~\cite{tay2020long}.

\begin{figure*}[tp]
\centering
\includegraphics[width=0.7\textwidth]{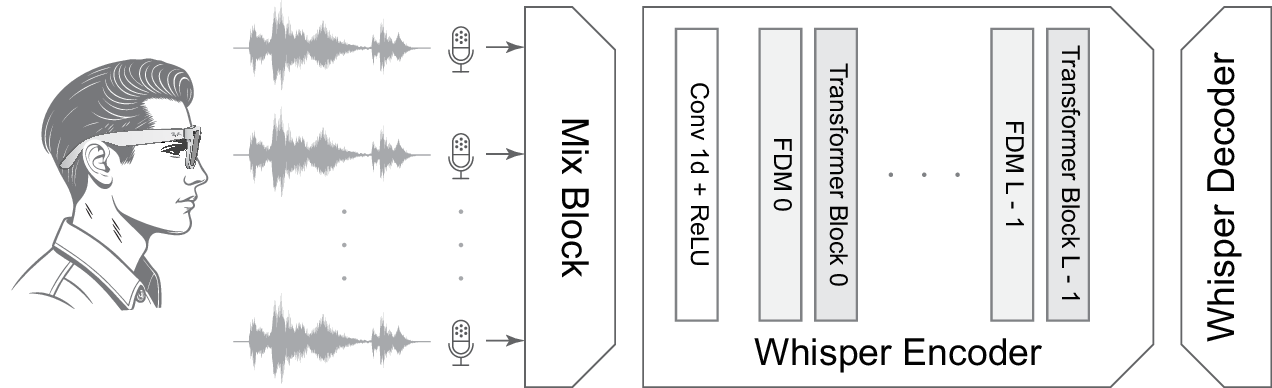}
\caption{Overall architecture of Side-Talk Rejection Multi-Microphone Whisper.}
\label{fig:flop1}
\end{figure*}

Mamba and Mamba-2~\cite{dao2024transformers} extend S4 for discrete data such as text and genomic sequences through two key innovations. First, they introduce an input-dependent selection mechanism that dynamically adjusts S4 parameters based on input data, unlike the time- and input-invariant S4. Second, Mamba incorporates a hardware-aware algorithm that scales linearly with sequence length, enhancing computational efficiency on modern hardware. By integrating S4 blocks with linear layers, Mamba achieves state-of-the-art performance across various long-sequence tasks, including text, speech, and genomics.

\section{Side-Talk Rejection Multi-Microphone Whisper}
\label{sec:method}

This section introduces side-talk rejection multi-microphone whisper (MMW), an extension of the Whisper architecture tailored for wearer-centric ASR on smart glasses. The proposed model leverages multi-channel audio waveforms to guide the transcription process, as illustrated in Fig.~\ref{fig:flop1}.

To enable wearer-aware processing without relying on explicit diarization labels or speaker embeddings, we infer frame-level speaker activity patterns directly from raw audio inputs. Let $S = \{s_0, s_1, \ldots, s_{n-1}\}$ denote the set of $n$ speakers, where $s_0$ is the target speaker (wearer). Let $C = \{c_0, c_1, \ldots, c_{m-1}\}$ represent the set of $m$ microphones available on the smart glasses and $X_C = \{x_0, x_1, \ldots, x_{m-1}\}$ represent the set of input wave.

\subsection{Mix Block}
OpenAI's Whisper model is originally designed to accept only single-channel log-mel spectrograms as input, thereby discarding valuable spatial cues that are critical for speaker separation. To retain spatial information, we utilize all available microphone channels $C$ from the smart glasses. Since the primary objective is to differentiate speakers rather than to explicitly localize them, we do not model the precise microphone array geometry. Instead, the model learns inter-microphone phase relationships directly from the raw waveform signals.

\begin{figure}[tp]
\centering
\includegraphics[width=0.38\textwidth]{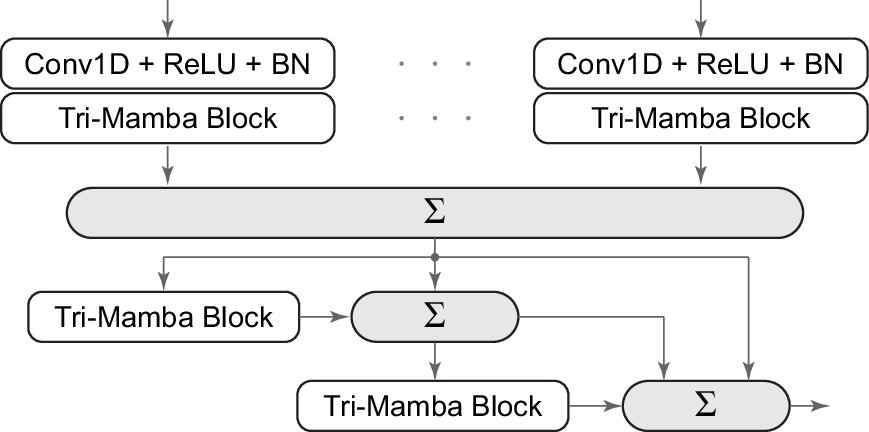}
\caption{Architecture of the Mix Block based on Tri-Mamba for multi-channel waveform fusion.}
\label{fig:flop2}
\end{figure}

The architecture of the proposed Mix Block is depicted in Fig.~\ref{fig:flop2}. To perform audio fusion, we employ a Tri-Mamba block \cite{liu2024masv}. The processing begins with a 1D convolutional layer, followed by ReLU activation and batch normalization (BN), which extracts low-level features from the multi-channel input. Subsequently, the features are passed through a Tri-Mamba block to capture bidirectional local temporal context.

Specifically, the forward hidden state at time step $t$ is updated according to:
\begin{equation}
\overrightarrow{\mathbf{A}}x(t) = \text{Mamba}(\overrightarrow{\mathbf{A}}x(t-1), \overrightarrow{\mathbf{c}}),
\end{equation}
where $\overrightarrow{\mathbf{c}}$ denotes the local context buffer in the forward direction. Similarly, the backward hidden state is computed as:
\begin{equation}
\overleftarrow{\mathbf{A}}x(t) = \text{Mamba}(\overleftarrow{\mathbf{A}}x(t+1), \overleftarrow{\mathbf{c}}).
\end{equation}

The final representation at each time step is obtained by concatenating the forward and backward hidden states:
\begin{equation}
\mathbf{A}x(t) = [\overrightarrow{\mathbf{A}}x(t); \overleftarrow{\mathbf{A}}x(t)].
\end{equation}


Finally, to improve feature preservation and facilitate gradient flow during training, full-scale skip connections are employed. These connections perform element-wise addition between the inputs and outputs of corresponding layers, helping to retain both low-level and high-level information and mitigating the risk of vanishing gradients in deeper architectures.

\subsection{Frame Diarization Mamba}

The Frame Diarization Mamba (FDM) layer is introduced to address three critical challenges. First, while Whisper was originally designed to process log-mel spectrogram inputs, the outputs of the Mix Block are audio embeddings that retain phase information, necessitating an adapter layer to bridge this mismatch. Second, introducing such an adapter allows efficient fine-tuning without requiring updates to Whisper’s core parameters, thereby preserving model stability. Third, the FDM layer acts as a masking mechanism, effectively suppressing bystander speech propagation and promoting the dominance of target speaker information in subsequent processing.

Let $\mathbf{Y}^l \in \mathbb{R}^{d_m \times T}$ denote the frame-wise input to the $l$-th transformer layer of Whisper. The FDM layer applies class-specific transformations to $\mathbf{Y}^l$ based on speaker activity patterns, resulting in the adapted frame representation $\tilde{\mathbf{y}}_t^l$ at each time step $t$:
\begin{equation}
\tilde{\mathbf{y}}_t^l = f_T(\mathbf{y}_t^l) p_T^t + f_{\mathcal{N}}(\mathbf{y}_t^l) p_{\mathcal{N}}^t + f_{\mathcal{O}}(\mathbf{y}_t^l) p_{\mathcal{O}}^t,
\end{equation}
where $p_T^t$, $p_{\mathcal{N}}^t$, and $p_{\mathcal{O}}^t$ denote the probabilities assigned to different speaker conditions: target-only ($\mathcal{T}$), non-target-only ($\mathcal{N}$), and overlapping speech ($\mathcal{O}$), respectively. Specifically, $\mathcal{T}$ refers to frames where only the target speaker $s_0$ is active, $\mathcal{N}$ corresponds to frames where one or more non-target speakers ($s \neq s_0$) are active while the target is silent, and $\mathcal{O}$ denotes frames where both the target and at least one non-target speaker are speaking simultaneously. Silent frames, where no speech is present, are treated as part of the target speaker class $\mathcal{T}$ to maintain continuity in the transcription process.

\subsection{Multi-Scale Group Relative Policy Optimization}

We extend Group Relative Policy Optimization (GRPO) to jointly optimize both audio enhancement quality and ASR transcription performance. Given a multi-channel input $X_C$, we sample $G$ enhanced outputs ${z_1, \ldots, z_G}$ from the enhancement policy $\pi_\theta(\cdot | X_C)$. Each output $z_j$ is evaluated using a composite reward function $r(X_C, z_j)$, which integrates three complementary objectives: frame-level consistency between the enhanced feature sequence and the ground-truth frame labels, utterance-level consistency between the dominant speaker condition inferred from the enhanced features and the ground-truth utterance labels, and ASR transcription quality measured by the word error rate (WER). These reward components together guide the optimization of the enhancement policy toward generating audio features that not only align with speaker activity patterns but also improve downstream transcription accuracy.

At each Transformer layer output $\mathbf{Y}^l \in \mathbb{R}^{d_m \times T}$, we predict the dominant speaker class for each frame by aggregating across the feature dimension $d_m$. To emphasize later frames—which are often more semantically informative—an exponential weighting $w_t$ is applied:
\begin{equation}
w_t = \exp\left( \beta \frac{t}{T} \right),
\end{equation}
where $\beta > 0$ is a scaling factor controlling the emphasis.
The predicted frame-wise dominant class sequence $\hat{y}_{1:T}$ is then compared with the ground-truth frame labels $y_{1:T}^\text{frame}$ using a normalized negative Hamming distance:
\begin{equation}
r_\text{frame}(X_C, z_j) = 1 - \frac{1}{\sum_{t=1}^T w_t} \sum_{t=1}^T w_t \cdot \mathbb{1}\left( \hat{y}_t \neq y_t^\text{frame} \right),
\end{equation}
where $\mathbb{1}(\cdot)$ is the indicator function.

Similarly, to compute utterance-level consistency, we first find the dominant frame class $\hat{y}\text{dom}$ across all frames:
\begin{equation}
\hat{y}\text{dom} = \arg\max_{k} \sum_{t=1}^T \mathbb{1}\left( \hat{y}_t = k \right),
\end{equation}
where $k$ indexes over possible speaker classes $\mathcal{T}, \mathcal{N}, \mathcal{O}$.

We then compare $\hat{y}\text{dom}$ to the ground-truth utterance label $y^\text{utt}$, with the reward defined as:
\begin{equation}
r\text{utt}(X_C, z_j) = \mathbb{1}\left( \hat{y}_\text{dom} = y^\text{utt} \right).
\end{equation}
Thus, the reward is 1 if the dominant speaker condition matches the ground-truth utterance label, and 0 otherwise.

Finally, we directly evaluate the ASR output quality. Let $\text{WER}(z_j)$ denote the word error rate of the ASR transcription generated from $z_j$. The reward is defined as:
\begin{equation}
r_\text{wer}(X_C, z_j) = 1 - \text{WER}(z_j),
\end{equation}
thus assigning higher rewards to lower error rates.

\vspace{2mm}

The overall composite reward is a weighted sum of the three components:
\begin{equation}
r(X_C, z_j) = \lambda_1 r_\text{frame}(X_C, z_j) + \lambda_2 r_\text{utt}(X_C, z_j) + \lambda_3 r_\text{wer}(X_C, z_j),
\end{equation}
where $\lambda_1$, $\lambda_2$, and $\lambda_3$ are hyperparameters controlling the relative importance of each reward term.

The group mean $\mu_{X_C}$ and standard deviation $\sigma_{X_C}$ of rewards are computed as:
\begin{align}
\mu_{X_C} &= \frac{1}{G} \sum_{j=1}^G r(X_C, z_j), \
\sigma_{X_C} &= \sqrt{ \frac{1}{G} \sum_{j=1}^G (r(X_C, z_j) - \mu_{X_C})^2 },
\end{align}
and the normalized advantage for each sample is:
\begin{equation}
\mathcal{A}(X_C, z_j) = \frac{r(X_C, z_j) - \mu_{X_C}}{\sigma_{X_C}}.
\end{equation}

The policy gradient update is given by:
\begin{equation}
\nabla_\theta \mathcal{J}(\theta) = \mathbb{E}{X_C \sim \mathcal{D}} \left[ \sum{j=1}^G \nabla_\theta \log \pi_\theta(z_j | X_C) \mathcal{A}(X_C, z_j) \right],
\end{equation}
where $\mathcal{D}$ denotes the training dataset.

\section{EXPERIMENTS}
\label{sec:typestyle}

\subsection{dataset}

Several spatial audio datasets have been introduced to support model development, including YouTube-360~\cite{morgado2020learning}, YouTube-ASMR~\cite{yang2020telling}, Pano-AVQA~\cite{yun2021pano}, and STARSS23~\cite{shimada2023starss23}. However, many of these datasets suffer from inconsistent quality and lack critical annotations, such as sound source direction or distance, limiting their applicability for spatial learning tasks.

We did not find any existing real or synthetic datasets specifically tailored for smart glasses scenarios, which typically involve wearable-centric speech capture under complex spatial conditions. To overcome the limitations of existing spatial datasets and better simulate smart glasses usage scenarios, we constructed a new real-world dataset using controlled playback and recording in an acoustic lab. The wearer was represented by a Sennheiser binaural dummy head fitted with Ray-Ban Meta smart glasses. These glasses include five microphones: four located on the arms (two on each side, positioned above and below), and a fifth microphone placed on the nose bridge. This configuration is designed to enhance voice clarity, suppress wind noise, and improve audio quality for both recording and voice interaction.

Speech signals were played from two directional speakers positioned around the dummy head to simulate bystander speakers, as shown in Fig. 3. The speakers were moved across multiple spatial configurations defined by four factors: azimuthal angle (eight discrete directions around the head), distance from the wearer (0.5 m or 2 m), vertical height offset ($\pm$0.5 m), and elevation angle ($\pm$45° tilt). These combinations allow us to capture diverse near-field and far-field spatial conditions relevant to wearable devices.

To further increase realism, background noise was introduced via wall-mounted speakers placed throughout the room. We used noise types consistent with the LibriMix framework, including CHiME-style ambient noise, music, and overlapping speech drawn from the MUSAN corpus. Noise levels were randomly varied between 30 dB and 90 dB SPL to simulate different environmental conditions. Finally, we collect 5k hours training set and 22 hours evaluation set. 

All speech content used in the playback was sampled from the LibriSpeech corpus and resampled to 16 kHz. This setup enables the construction of a rich and controlled spatial dataset with fine-grained control over directionality, distance, and interference—providing a strong foundation for evaluating smart-glasses-based speech processing models.

\begin{figure}[htb]

\begin{minipage}[b]{1.0\linewidth}
  \centering
  \centerline{\includegraphics[width=8.5cm]{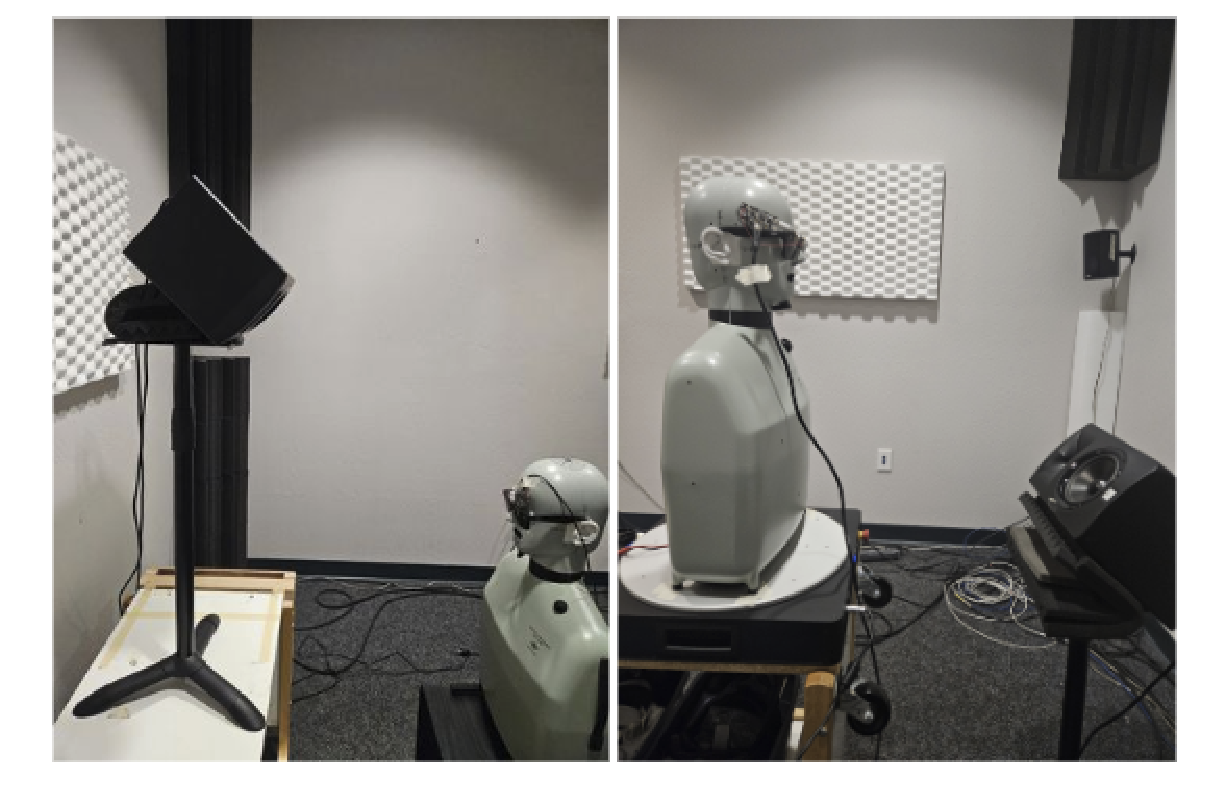}}
  \subcaption{Physical Test Environment: Microphone and Speaker Arrangement}
\end{minipage}
\begin{minipage}[b]{1.0\linewidth}
  \centering
  \centerline{\includegraphics[width=8.5cm]{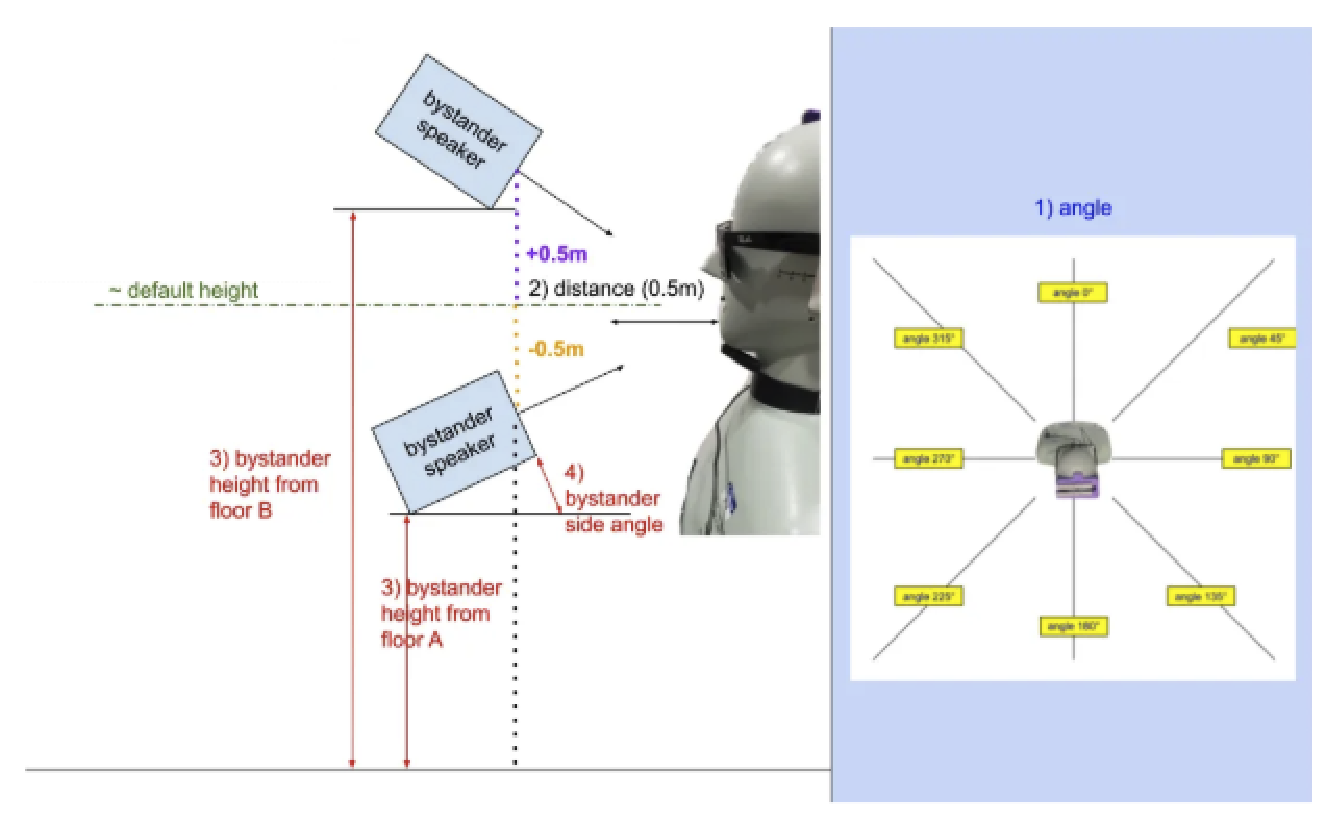}}
  \subcaption{Variable Configuration Diagram: Angle, Distance, and Height Parameters for Bystander Speaker Placement}
\end{minipage}
\caption{Setup and Parameter Definition for Bystander Speaker Experiments}
\label{fig:res}
\end{figure}

\subsection{Training details}
To improve the performance of Whisper, we applied architectural enhancements targeting memory and efficiency constraints arising from its large vocabulary size (50k) and fixed sequence length. We conducted ablation studies using Whisper-medium.en, whereas the final model was trained with Whisper-large-v3-turbo. All training was carried out using bf16 precision and the AdamW optimizer \cite{loshchilov2017decoupled}, with a batch size of 64. A linear decay schedule was employed, with 2,000 warm-up steps. Subsequently, the entire model was fine-tuned for up to 50,000 steps, with early stopping activated after five epochs without improvement. Most models achieved convergence within ten epochs. Final evaluation metrics were based on the development set, selecting the checkpoint that achieved the lowest word error rate (WER).

\section{Ablation Study}
\label{sec:ablation}

We conduct an ablation study to evaluate the contribution of each proposed component, as shown in Table~\ref{tab:ablation}. We first directly evaluate the original Whisper model on our multi-microphone dataset, achieving a Word Error Rate of 67.81\%. The high error rate indicates that the unmodified Whisper model transcribes both wearer and bystander speech without distinction. Introducing the FDM layer improves WER to 32.61\%. FDM focuses the model's attention on wearer-dominant frames by masking bystander features. However, because FDM does not leverage explicit bio-information or spatial priors, the improvement is limited. Next, incorporating the GRPO strategy further reduces WER. Using GRPO with only WER loss achieves 28.42\%. However, when combining frame-level and utterance-level rewards, WER drops substantially to 13.10\%. This demonstrates the importance of utterance-level guidance: frame-level labels can be noisy due to silence segments (e.g., recording gaps), whereas utterance-level supervision provides a cleaner and faster learning signal. Finally, replacing the traditional beamforming + log-mel frontend with our proposed Mix Block yields a final WER of 3.71\%. Mix Block preserves phase information and spatial cues that are lost in conventional log-mel extraction, thereby significantly enhancing performance.

\begin{table}[ht]
\centering
\caption{Ablation study results on evaluation set.}
\label{tab:ablation}
\begin{tabular}{lc}
\toprule
\textbf{Method} & \textbf{WER (\%)} \\
\midrule
Whisper (direct eval) & 67.81 \\
+ FDM & 32.61 \\
+ GRPO (WER only) & 28.42 \\
+ GRPO (Frame + Utterance + WER) & 13.10 \\
+ Mix Block & 3.71 \\
\bottomrule
\end{tabular}
\end{table}
\subsection{Comparison to Baselines}

\begin{table}[ht]
\centering
\caption{Comparison of TS-ASR Whisper and LLM-based models. Results for Vicuna and Llama 4 are based on direct fine-tuning.}
\vspace{0.5em}
\begin{tabular}{lc}
\toprule
\textbf{System} & \textbf{WER (\%)} \\
\midrule
\multicolumn{2}{l}{\textit{No ground truth segmentation}} \\
TS-ASR Whisper & \textbf{14.9} \\
MMW Whisper  & \textbf{12.7} \\
\midrule
\multicolumn{2}{l}{\textit{With ground truth segmentation}} \\
Input masking & 56.7 \\
Ma et al. \cite{ma2024extending} & 26.4 \\
Zhang et al. \cite{zhang2023weakly} & 23.5 \\
TS-ASR Whisper & \textbf{10.9} \\
MMW Whisper  & \textbf{8.3} \\
\midrule
\multicolumn{2}{l}{\textit{Fine-tuning with LLMs}} \\
Vicuna (7B) & 12.8 \\
Llama 4 (8B) & 9.33 \\
MMW + Llama 4 & \textbf{4.53} \\
\bottomrule
\end{tabular}
\label{tab:libri2mix_comparison}
\end{table}

Table~\ref{tab:libri2mix_comparison} summarizes the performance of our proposed MMW Whisper system alongside various baseline models and fine-tuned large language models (LLMs) on the Libri2Mix test-both set. For fair comparison, we only use single input for MMW Whisper for "No ground truth segmentation" and "With ground truth segmentation". 

In the category without access to ground-truth segmentation, MMW Whisper achieves a WER of 12.7\%, substantially outperforming earlier end-to-end systems such as Raj et al.\cite{raj2023surt} and Vinnikov et al.\cite{vinnikov2024notsofar}. When ground-truth segmentation is provided, MMW Whisper further improves to a WER of 8.3\%, outperforming both the naive input masking baseline (56.7\%) and more advanced modular systems such as Ma et al.\cite{ma2024extending} (26.4\%) and Zhang et al.\cite{zhang2023weakly} (23.5\%). These results demonstrate that while segmentation information provides a strong supervisory signal, our architecture is capable of leveraging it more effectively than conventional diarization-plus-ASR pipelines. The robustness of MMW Whisper under both supervised and unsupervised conditions underlines its flexibility and generalization capabilities.

Beyond modular and end-to-end baselines, we also explored the fine-tuning of large language models directly for the multi-talker ASR task. Fine-tuned Vicuna (7B) and Llama 4 (8B) achieve WERs of 12.8\% and 9.33\%, respectively, validating that modern LLMs possess strong transferability and adaptation abilities, even without architectural specialization for speech processing. Despite the strong results from LLM fine-tuning, our proposed MMW + Llama 4 system achieves a WER of 4.53\%. This highlights the benefit of integrating spatial-aware multi-microphone features through the Mix Block, allowing the model to retain fine-grained directional and phase information that traditional log-mel pipelines discard.

Overall, the results indicate three key observations. First, fine-tuned LLMs, even without access to multi-microphone information, can outperform traditional ASR systems under overlapping conditions. Second, preserving multi-microphone spatial information is critical for further improving recognition accuracy in complex acoustic environments. Third, MMW Whisper shows strong performance across both supervised and unsupervised settings, demonstrating its ability to implicitly learn wearer-centric speech features without heavy reliance on external annotations. These findings collectively push the boundaries of multi-speaker ASR performance, offering promising directions for future research.

\section{Conclusion}
\label{sec:conclusion}

We proposed Side-Talk Rejection Multi-Microphone Whisper (MMW), a new framework that enhances wearer-centric ASR on smart glasses by integrating a Mix Block for multi-channel raw waveform fusion, a Frame Diarization Mamba Layer for side-talk suppression, and a Multi-Scale Group Relative Policy Optimization (GRPO) strategy. Experiments on a real-world multi-microphone dataset demonstrated that MMW significantly reduces WER compared to both traditional modular systems and fine-tuned large language models, achieving a WER of 3.71\% in ablation studies and 4.95\% when combined with Llama 4. These results validate the effectiveness of preserving spatial cues at the waveform level and optimizing both frame- and utterance-level objectives for robust multi-speaker ASR in real-world environments.


\bibliographystyle{IEEEbib}
\bibliography{strings,refs}

\begin{thebibliography}{10}

\bibitem{achiam2023gpt}
Josh Achiam, Steven Adler, Sandhini Agarwal, Lama Ahmad, Ilge Akkaya,
  Florencia~Leoni Aleman, Diogo Almeida, Janko Altenschmidt, Sam Altman,
  Shyamal Anadkat, et~al.,
\newblock ``Gpt-4 technical report,''
\newblock {\em arXiv preprint arXiv:2303.08774}, 2023.

\bibitem{touvron2023llama}
Hugo Touvron, Thibaut Lavril, Gautier Izacard, Xavier Martinet, Marie-Anne
  Lachaux, Timoth{\'e}e Lacroix, Baptiste Rozi{\`e}re, Naman Goyal, Eric
  Hambro, Faisal Azhar, et~al.,
\newblock ``Llama: Open and efficient foundation language models,''
\newblock {\em arXiv preprint arXiv:2302.13971}, 2023.

\bibitem{van1988beamforming}
Barry~D Van~Veen and Kevin~M Buckley,
\newblock ``Beamforming: A versatile approach to spatial filtering,''
\newblock {\em IEEE assp magazine}, vol. 5, no. 2, pp. 4--24, 1988.

\bibitem{knapp2003generalized}
Charles Knapp and Glifford Carter,
\newblock ``The generalized correlation method for estimation of time delay,''
\newblock {\em IEEE transactions on acoustics, speech, and signal processing},
  vol. 24, no. 4, pp. 320--327, 2003.

\bibitem{schmidt1986multiple}
Ralph Schmidt,
\newblock ``Multiple emitter location and signal parameter estimation,''
\newblock {\em IEEE transactions on antennas and propagation}, vol. 34, no. 3,
  pp. 276--280, 1986.

\bibitem{roy1989esprit}
Richard Roy and Thomas Kailath,
\newblock ``Esprit-estimation of signal parameters via rotational invariance
  techniques,''
\newblock {\em IEEE Transactions on acoustics, speech, and signal processing},
  vol. 37, no. 7, pp. 984--995, 1989.

\bibitem{adavanne2018sound}
Sharath Adavanne, Archontis Politis, Joonas Nikunen, and Tuomas Virtanen,
\newblock ``Sound event localization and detection of overlapping sources using
  convolutional recurrent neural networks,''
\newblock {\em IEEE Journal of Selected Topics in Signal Processing}, vol. 13,
  no. 1, pp. 34--48, 2018.

\bibitem{xu2017convolutional}
Yong Xu, Qiuqiang Kong, Qiang Huang, Wenwu Wang, and Mark~D Plumbley,
\newblock ``Convolutional gated recurrent neural network incorporating spatial
  features for audio tagging,''
\newblock in {\em 2017 International Joint Conference on Neural Networks
  (IJCNN)}. IEEE, 2017, pp. 3461--3466.

\bibitem{park2021many}
Sooyoung Park, Youngho Jeong, and Taejin Lee,
\newblock ``Many-to-many audio spectrogram tansformer: Transformer for sound
  event localization and detection.,''
\newblock in {\em DCASE}, 2021, pp. 105--109.

\bibitem{zheng2024bat}
Zhisheng Zheng, Puyuan Peng, Ziyang Ma, Xie Chen, Eunsol Choi, and David
  Harwath,
\newblock ``Bat: Learning to reason about spatial sounds with large language
  models,''
\newblock {\em arXiv preprint arXiv:2402.01591}, 2024.

\bibitem{tang2024can}
Changli Tang, Wenyi Yu, Guangzhi Sun, Xianzhao Chen, Tian Tan, Wei Li, Jun
  Zhang, Lu~Lu, Zejun Ma, Yuxuan Wang, et~al.,
\newblock ``Can large language models understand spatial audio?,''
\newblock {\em arXiv preprint arXiv:2406.07914}, 2024.

\bibitem{chen2022soundspaces}
Changan Chen, Carl Schissler, Sanchit Garg, Philip Kobernik, Alexander Clegg,
  Paul Calamia, Dhruv Batra, Philip Robinson, and Kristen Grauman,
\newblock ``Soundspaces 2.0: A simulation platform for visual-acoustic
  learning,''
\newblock {\em Advances in Neural Information Processing Systems}, vol. 35, pp.
  8896--8911, 2022.

\bibitem{yang2022srp}
Bing Yang, Hong Liu, and Xiaofei Li,
\newblock ``Srp-dnn: Learning direct-path phase difference for multiple moving
  sound source localization,''
\newblock in {\em ICASSP 2022-2022 IEEE International Conference on Acoustics,
  Speech and Signal Processing (ICASSP)}. IEEE, 2022, pp. 721--725.

\bibitem{qiao2024joint}
Minglang Qiao, Yufan Liu, Mai Xu, Xin Deng, Bing Li, Weiming Hu, and Ali Borji,
\newblock ``Joint learning of audio--visual saliency prediction and sound
  source localization on multi-face videos,''
\newblock {\em International Journal of Computer Vision}, vol. 132, no. 6, pp.
  2003--2025, 2024.

\bibitem{devnani2024learning}
Bhavika Devnani, Skyler Seto, Zakaria Aldeneh, Alessandro Toso, Elena
  Menyaylenko, Barry-John Theobald, Jonathan Sheaffer, and Miguel Sarabia,
\newblock ``Learning spatially-aware language and audio embeddings,''
\newblock {\em Advances in Neural Information Processing Systems}, vol. 37, pp.
  33505--33537, 2024.

\bibitem{tang2023salmonn}
Changli Tang, Wenyi Yu, Guangzhi Sun, Xianzhao Chen, Tian Tan, Wei Li, Lu~Lu,
  Zejun Ma, and Chao Zhang,
\newblock ``Salmonn: Towards generic hearing abilities for large language
  models,''
\newblock {\em arXiv preprint arXiv:2310.13289}, 2023.

\bibitem{ghosh2024gama}
Sreyan Ghosh, Sonal Kumar, Ashish Seth, Chandra Kiran~Reddy Evuru, Utkarsh
  Tyagi, S~Sakshi, Oriol Nieto, Ramani Duraiswami, and Dinesh Manocha,
\newblock ``Gama: A large audio-language model with advanced audio
  understanding and complex reasoning abilities,''
\newblock {\em arXiv preprint arXiv:2406.11768}, 2024.

\bibitem{hu2022lora}
Edward~J Hu, Yelong Shen, Phillip Wallis, Zeyuan Allen-Zhu, Yuanzhi Li, Shean
  Wang, Lu~Wang, Weizhu Chen, et~al.,
\newblock ``Lora: Low-rank adaptation of large language models.,''
\newblock {\em ICLR}, vol. 1, no. 2, pp. 3, 2022.

\bibitem{radford2023robust}
Alec Radford, Jong~Wook Kim, Tao Xu, Greg Brockman, Christine McLeavey, and
  Ilya Sutskever,
\newblock ``Robust speech recognition via large-scale weak supervision,''
\newblock in {\em International conference on machine learning}. PMLR, 2023,
  pp. 28492--28518.

\bibitem{ma2024extending}
Hao Ma, Zhiyuan Peng, Mingjie Shao, Jing Li, and Ju~Liu,
\newblock ``Extending whisper with prompt tuning to target-speaker asr,''
\newblock in {\em ICASSP 2024-2024 IEEE International Conference on Acoustics,
  Speech and Signal Processing (ICASSP)}. IEEE, 2024, pp. 12516--12520.

\bibitem{meng2024empowering}
Lingwei Meng, Jiawen Kang, Yuejiao Wang, Zengrui Jin, Xixin Wu, Xunying Liu,
  and Helen Meng,
\newblock ``Empowering whisper as a joint multi-talker and target-talker speech
  recognition system,''
\newblock {\em arXiv preprint arXiv:2407.09817}, 2024.

\bibitem{polok2024dicow}
Alexander Polok, Dominik Klement, Martin Kocour, Jiangyu Han, Federico Landini,
  Bolaji Yusuf, Matthew Wiesner, Sanjeev Khudanpur, Jan {\v{C}}ernock{\`y}, and
  Luk{\'a}{\v{s}} Burget,
\newblock ``Dicow: Diarization-conditioned whisper for target speaker automatic
  speech recognition,''
\newblock {\em arXiv preprint arXiv:2501.00114}, 2024.

\bibitem{polok2025target}
Alexander Polok, Dominik Klement, Matthew Wiesner, Sanjeev Khudanpur, Jan
  {\v{C}}ernock{\`y}, and Luk{\'a}{\v{s}} Burget,
\newblock ``Target speaker asr with whisper,''
\newblock in {\em ICASSP 2025-2025 IEEE International Conference on Acoustics,
  Speech and Signal Processing (ICASSP)}. IEEE, 2025, pp. 1--5.

\bibitem{liu2024masv}
Yang Liu, Li~Wan, Yiteng Huang, Ming Sun, Yangyang Shi, and Florian Metze,
\newblock ``Masv: Speaker verification with global and local context mamba,''
\newblock {\em arXiv preprint arXiv:2412.10989}, 2024.

\bibitem{liu2024exploration}
Yunpeng Liu, Xukui Yang, and Dan Qu,
\newblock ``Exploration of whisper fine-tuning strategies for low-resource
  asr,''
\newblock {\em EURASIP Journal on Audio, Speech, and Music Processing}, vol.
  2024, no. 1, pp. 29, 2024.

\bibitem{rouditchenko2024whisper}
Andrew Rouditchenko, Yuan Gong, Samuel Thomas, Leonid Karlinsky, Hilde Kuehne,
  Rogerio Feris, and James Glass,
\newblock ``Whisper-flamingo: Integrating visual features into whisper for
  audio-visual speech recognition and translation,''
\newblock {\em arXiv preprint arXiv:2406.10082}, 2024.

\bibitem{gu2021efficiently}
Albert Gu, Karan Goel, and Christopher R{\'e},
\newblock ``Efficiently modeling long sequences with structured state spaces,''
\newblock {\em arXiv preprint arXiv:2111.00396}, 2021.

\bibitem{gu2020hippo}
Albert Gu, Tri Dao, Stefano Ermon, Atri Rudra, and Christopher R{\'e},
\newblock ``Hippo: Recurrent memory with optimal polynomial projections,''
\newblock {\em Advances in neural information processing systems}, vol. 33, pp.
  1474--1487, 2020.

\bibitem{vaswani2017attention}
Ashish Vaswani, Noam Shazeer, Niki Parmar, Jakob Uszkoreit, Llion Jones,
  Aidan~N Gomez, {\L}ukasz Kaiser, and Illia Polosukhin,
\newblock ``Attention is all you need,''
\newblock {\em Advances in neural information processing systems}, vol. 30,
  2017.

\bibitem{tay2020long}
Yi~Tay, Mostafa Dehghani, Samira Abnar, Yikang Shen, Dara Bahri, Philip Pham,
  Jinfeng Rao, Liu Yang, Sebastian Ruder, and Donald Metzler,
\newblock ``Long range arena: A benchmark for efficient transformers,''
\newblock {\em arXiv preprint arXiv:2011.04006}, 2020.

\bibitem{dao2024transformers}
Tri Dao and Albert Gu,
\newblock ``Transformers are ssms: Generalized models and efficient algorithms
  through structured state space duality,''
\newblock {\em arXiv preprint arXiv:2405.21060}, 2024.

\bibitem{morgado2020learning}
Pedro Morgado, Yi~Li, and Nuno Nvasconcelos,
\newblock ``Learning representations from audio-visual spatial alignment,''
\newblock {\em Advances in Neural Information Processing Systems}, vol. 33, pp.
  4733--4744, 2020.

\bibitem{yang2020telling}
Karren Yang, Bryan Russell, and Justin Salamon,
\newblock ``Telling left from right: Learning spatial correspondence of sight
  and sound,''
\newblock in {\em Proceedings of the IEEE/CVF conference on computer vision and
  pattern recognition}, 2020, pp. 9932--9941.

\bibitem{yun2021pano}
Heeseung Yun, Youngjae Yu, Wonsuk Yang, Kangil Lee, and Gunhee Kim,
\newblock ``Pano-avqa: Grounded audio-visual question answering on 360deg
  videos,''
\newblock in {\em Proceedings of the IEEE/CVF International Conference on
  Computer Vision}, 2021, pp. 2031--2041.

\bibitem{shimada2023starss23}
Kazuki Shimada, Archontis Politis, Parthasaarathy Sudarsanam, Daniel~A Krause,
  Kengo Uchida, Sharath Adavanne, Aapo Hakala, Yuichiro Koyama, Naoya
  Takahashi, Shusuke Takahashi, et~al.,
\newblock ``Starss23: An audio-visual dataset of spatial recordings of real
  scenes with spatiotemporal annotations of sound events,''
\newblock {\em Advances in neural information processing systems}, vol. 36, pp.
  72931--72957, 2023.

\bibitem{loshchilov2017decoupled}
Ilya Loshchilov and Frank Hutter,
\newblock ``Decoupled weight decay regularization,''
\newblock {\em arXiv preprint arXiv:1711.05101}, 2017.

\bibitem{zhang2023weakly}
Wangyou Zhang and Yanmin Qian,
\newblock ``Weakly-supervised speech pre-training: A case study on target
  speech recognition,''
\newblock {\em arXiv preprint arXiv:2305.16286}, 2023.

\bibitem{raj2023surt}
Desh Raj, Daniel Povey, and Sanjeev Khudanpur,
\newblock ``Surt 2.0: Advances in transducer-based multi-talker speech
  recognition,''
\newblock {\em IEEE/ACM Transactions on Audio, Speech, and Language
  Processing}, vol. 31, pp. 3800--3813, 2023.

\bibitem{vinnikov2024notsofar}
Alon Vinnikov, Amir Ivry, Aviv Hurvitz, Igor Abramovski, Sharon Koubi, Ilya
  Gurvich, Shai Peer, Xiong Xiao, Benjamin~Martinez Elizalde, Naoyuki Kanda,
  et~al.,
\newblock ``Notsofar-1 challenge: New datasets, baseline, and tasks for distant
  meeting transcription,''
\newblock {\em arXiv preprint arXiv:2401.08887}, 2024.

\end{thebibliography}

\end{document}